\documentclass[preprintnumbers,amsmath,amssymb,twocolumn]{revtex4}
\usepackage[colorlinks=true,citecolor=red,filecolor=green,linkcolor=blue,pdfnewwindow=true]{hyperref}
\usepackage{makeidx}
\usepackage{bm}
\usepackage[title]{appendix}

\usepackage{graphicx}
\usepackage{epstopdf}
\usepackage{amsmath} % For typesetting math
\usepackage{amssymb} % Adds new symbols to be used in math mode
\usepackage{booktabs} % Top and bottom rules for tables
\usepackage{enumitem} % Used to reduce itemize/enumerate spacing
\usepackage{palatino} % Use the Palatino font
\usepackage[svgnames]{xcolor} % Specify colors by their 'svgnames', for a full list of all colors available see here: http://www.latextemplates.com/svgnames-colors

\def\prl#1#2#3{{ Phys. Rev. Lett.} {\bf #1}, #2 (#3)}
\def\pre#1#2#3{{ Phys. Rev. E.} {\bf #1}, #2 (#3)}
\def\pnas#1#2#3{{ PNAS. } {\bf #1}, #2 (#3)}

\def\ijbc#1#2#3{International Journal of Bifurcation and Chaos {\bf #1}, #2 (#3)}

\def\pla#1#2#3{Physics Letters A {\bf#1},  #2 {(#3)}}

\def\epj#1#2#3{Eur. Phys. J. {\bf #1}, #2 (#3)}

\def\chaos#1#2#3{Chaos {\bf #1}, #2 (#3)}

\def\ep{\varepsilon}

\def\ie{i.e. }

\def\beqr{\begin{eqnarray}}
\def\eqnr{\end{eqnarray}}
\def\beq{\begin{equation}}
\def\bc{\begin{center}}
\def\ec{\end{center}}
\def\eqn{\end{equation}\noindent}
\begin{document}
\title{Controlling chimera states in multilayer network through linear augmentation}
\author{Anjuman Ara Khatun and Haider Hasan Jafri}
\affiliation{Department of Physics, Aligarh Muslim University, Aligarh 202 002, India}
%\date{\today}
\begin{abstract}

In this work we present a method to control chimera states through linear augmentation. Using an ensemble of globally coupled oscillators, we demonstrate that control over the spatial location of the incoherent or coherent regions of a chimera state in a network can be achieved. This is characterized by exploring their basins of attraction. We also verify the stability of different states present in the coupled systems before and after linear augmentation by calculating the transverse Lyapunov exponent and Master stability function. We observe that the control through this technique is independent of the coupling mechanisms and also on the initial conditions chosen for the creation of chimera states.
\end{abstract}
\maketitle
\begin{section}{Introduction}

Chimera states are interesting collective behavior where synchronized and desynchronized dynamics coexists. These are observed in an ensemble of either globally or nonlocally coupled oscillators. In classical settings, a group of nonlocally coupled identical phase oscillators can spontaneously form two groups. One that posses synchronized motion, while the other group is desynchronized \cite{kuramoto, strogatz}. This coexistence of coherent and incoherent group was called as chimera state. Numerous studies have been devoted to study the chimera states in various generalized situations. These include ensembles of metronomes \cite{thutupalli,kapitaniak}, networks in neuroscience \cite{battagila, vicente}, biology \cite{rattenborg,laing}, Josephson junction arrays \cite{wiesenfield} and electrochemical systems \cite{mazouz,garcia}. Recently, there have been studies where chimeric states were observed as a result of induced multistability in identical chaotic oscillators that have been coupled without time delay \cite{chandrasekar,sangeeta-pre,anjuman}. Multistability is induced in the system as a result of coupling where the basins of the coexisting attractors are completely intertwined leading to stable chimera states for arbitrary initial conditions. Transitions in these intertwined or riddled basins \cite{yclai} have been characterized by looking at the transverse Lyapunov exponents, uncertainty exponent and scaling laws \cite{viana, sangeeta-chaos}.

Controlling chimera states has been an important concern in the field of applied complex systems. Some studies have shown that chimeric states can be stabilized in nonlocally coupled oscillator networks. Gradient dynamics was applied to control chimera states, resulting in chimeras with desired positions of coherent and incoherent domains \cite{bick}. A feedback control \cite{sieber} scheme was proposed to control the basins of attraction and lifetime of chimeras in large coupled systems. Chimera states have also been controlled by suppressing the chimera collapse and stabilizing its spatial positions \cite{omel16}. Chimera states in a small network was controlled by making use of tweezers \cite{omel18,omel19}. At other instance, chimera states were controlled with the help of delays where life time of the amplitude chimeras can be greatly increased or decreased depending upon the coupling delay \cite{gjurchinovski}. Moreover, a self feedback can be applied to control the spatial positions of coherent and incoherent domains in the array of coupled oscillators \cite{bera-control}.

Recently, it has been shown that linear augmentation (LA) has powerful effects in targeting desired fixed-point solutions \cite{sha}, suppressing bistability \cite{shr}, controlling the dynamics of drive-response systems \cite{sin} and regulating the dynamics of hidden attractors \cite{pra}. The advantage of this control scheme is that it allows to achieve the desired state without manipulating
system parameters. Here we implement LA technique to control chimera states. These chimera states may emerge through different types of couplings namely global, mean-field, nonlocal and local. The idea of LA has not been previously explored to control the chimera states.   

In this work we consider a network of $N$ identical chaotic oscillators coupled to a common drive. The dynamics of each oscillator evolves according to the rule $\dot{\overrightarrow{\bf {X}}} = F(\overrightarrow{\bf{X}}) ~ (\overrightarrow{\bf{X}} \in \mathbb{R}^m$). Out of the $m$ state variables of each dynamical system one is connected to a common drive. We argue that our system can be represented as a multilayer network with $l = m + 2$ layers. The multilayer network consists of three layers corresponding to the variables of response systems \cite{sevil}, a layer consisting of driving nodes and a control layer where each node is a linear system. The interlayer connections \cite{bocca, kivela} are governed by the dynamics of the Lorenz system and its connection to a common drive as argued in the following section. If all the nodes of one of the response layers are coupled to a common drive, one observes chimera states through induced multistability \cite{sangeeta-pre}. Chimeras may be controlled by augmenting the desired nodes of the response layers to that of the control layer. The resulting state may be a chimera where coherent and incoherent domains have fixed spatial locations. By changing the number of augmented nodes one can also obtain a state that is purely synchronized, coexisting clusters or a desynchronized state. The underlying mechanism behind the annihilation of coexisting attractors can be understood by exploring the basins of attraction of the linearly augmented response system driven by a chaotic system. We observe that in the absence of linear augmentation, the basin is riddled having multiple attractors. For a proper choice of control parameter riddling disappears because the response has been stabilized to one of the desired chaotic attractor. The stability and robustness of these resulting states can be verified with the help of  Master Stability function \cite{pecora-msf,huang, bocaletti-msf} which is negative for the stable synchronized state. 
 
We describe the creation of chimera states in the framework of multilayer networks in Section \ref{cres}. This is then extended to study the control of chimera states in Section \ref{clinaug}. Further, in section \ref{control} we explore the parameter space to characterize regions where effective control is permissible. In Section \ref{robust} we explore the robustness and stability of the resulting states. This is followed by a summary and discussion in Section \ref{conclusion}. 
\end{section}

\section{Collective dynamics of the network of response systems}
\label{cres}
\begin{figure}[htb]  
\centering
\scalebox{0.28}{\includegraphics{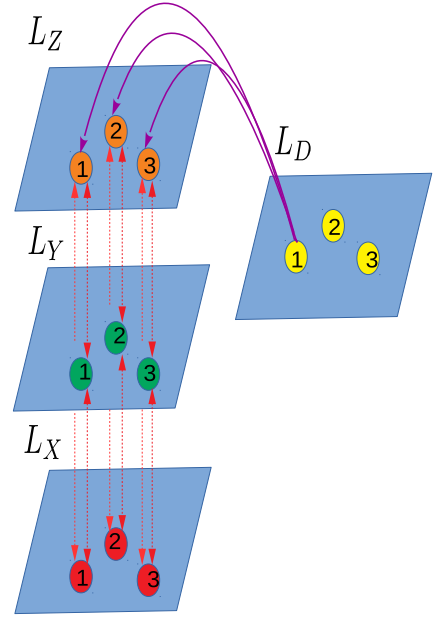}}
\caption{Multilayer diagram of three response systems whose $X_r$, $Y_r$, $Z_r$ variables are the nodes of $L_X $, $L_Y$, $L_Z$ layers, respectively. $X_r$ variables of response systems are driven by a single node of the $L_D$ layer. Magenta color (dark gray) curves represent real physical couplings between a drive and response systems. Bidirectional dotted lines represent explicit dependence of variables of each response system.}
\label{fig0}
\end{figure}

In the present work we create chimera state in a network of $N$ coupled chaotic Lorentz systems $\dot{\overrightarrow{\bf{X}}}_r = F(\overrightarrow{\bf{X}}_r)$ driven by a chaotic R\"ossler drive $\dot{\overrightarrow{\bf{X}}}_d = F(\overrightarrow{\bf{X}}_d)$, $( \overrightarrow{\bf{X}}_r, \overrightarrow{\bf{X}}_d \in {\mathbb{R}^m})$. In this case for the sake of simplicity, we present coupled systems as a network consisting of $l = m + 1$ layers as shown in Fig.~\ref{fig0}. The dynamics of each node in layers $L_X$, $L_Y$, $L_Z$ is given by the three variables $ X_r, Y_r, Z_r$, respectively of the Lorenz (response)
systems \cite{sevil,bocca,kivela,gamb,leyva,long}. Dynamics at each node of the $L_D$ layer is described by the R\"ossler drive. Only one node in the $L_D$ layer is used to drive the entire system. Other nodes in the $L_D$ layer remains disconnected from the network. The real physical couplings present between the nodes of layers $L_Z$ and $L_D$ is explained schematically in Fig.~\ref{fig0} \cite{sevil}. Since there is no real coupling between the nodes in $L_X$ and $L_Y$ layer we may called them as improper layers of the network \cite{sevil}. The connection topology
between the nodes of each layer (Fig.~\ref{fig0}) can be represented by an adjacency matrix $A_{ij}^{[\alpha]}$, 
where $\alpha = L_{X}, L_{Y}, L_{Z}, L_{D}$ identifies different layers and $i,j = 1, 2, 3, ...., N $ represent different nodes of the layer present in the network. $A_{ij}^{[\alpha]} = 1$, if there is a connection between the nodes, otherwise it is zero \cite{bocca, kivela,fede}. 
In layer $L_{Z}$, each nodes is connected to only one node of the $L_D$ layer \ie layer $L_{Z}$ and $L_D$ makes interlayer 
connection \cite{bocca, kivela} between them. The elements of the matrix $A_{ij}^{[\alpha]} = A_{i1}^{[L_{Z}L_{D}]} = A_{i1} = 1$ for 
all $i = 1, 2, 3, ...., N$ and $j = 1$ (only one node of the $L_D$ layer is connected). For each layer $A_{ij}^{[\alpha]} = 0 $ if there is no intralayer connection between the nodes of a particular layer \cite{bocca, kivela}. We present explicitly the dependence of the state variables of each response systems by the bidirectional dotted lines.

Mathematically, an ensemble of $N$ Lorenz oscillators driven by a chaotic drive namely R\"ossler oscillator may be described by the following equations
\begin{eqnarray}
\label{eq1}
\dot{x_d}&=& -{y_d} -{z_d}\nonumber \\ 
\dot{y_d}&=& {x_d}+a{y_d}\nonumber \\ 
\dot{z_d}&=&b+{z_d}({x_d}-c) \nonumber \\
\dot{x}_{r_i}&=& \sigma(y_{r_i}-x_{r_i}) \nonumber \\
\dot{y}_{r_i}&=& rx_{r_i} - y_{r_i} - x_{r_i}z_{r_i} \nonumber \\
\dot{z}_{r_i}&=& x_{r_i}y_{r_i} - \beta z_{r_i} + \ep_1 A_{ij}\left(z_d - z_{r_i}\right) ,
\end{eqnarray}
where, subscripts $d$ and $r$ represent drive and response systems, respectively.  $i = 1,2,3,....,N$ are the number of response systems that represent each node of the layers $L_X,L_Y$ and $L_Z$. Since only one drive is used to study the dynamics of Eq.~\ref{eq1}, therefore $j = 1$ \ie  $A_{ij} = A_{i1}$. $\ep_1$ is the coupling strength between drive and response systems. The parameter values are taken to be $a = 0.2$, $b = 0.2$, $c = 5.7$, $\sigma = 10$, $r = 28$ and $\beta = 8/3$ such that the dynamics of the uncoupled systems is chaotic. Since the Lorentz system is invariant under the transformation $(-x_r,-y_r,z_r) \rightarrow  (x_r,y_r,z_r)$, we couple the system in $z_r$ variable thereby preserving the symmetry of the system in the $x_r-y_r$  plane \cite{sangeeta-chaos}. $A_{i1}$ is an $(N \times 1)$ adjacency matrix given by
$\begin{bmatrix}
1 & \dots & 1 
\end{bmatrix}^T$
\quad
where, $T$ is the transpose. This represents interlayer connection between the nodes of $L_Z$ and $L_D$ layers. 

\begin{figure}[htp]  
\centering
\scalebox{0.45}{\includegraphics{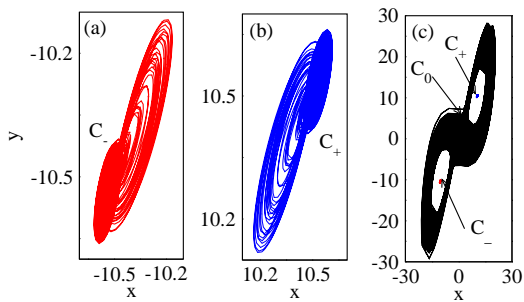}}
\caption{Attractors (a) $C_-$, (b) $C_+$ and (c) $C_0$ are created due to multistability in Eq.~\ref{eq1} with $N = 1$ at coupling strength $\ep_1 = 0.70$.}
\label{fig1}
\end{figure}

For $N = 1$, Eq.~\ref{eq1} reduces to the drive response configuration which has been studied in Ref.~\cite{sangeeta-chaos}. The main observations can be summarized as follows. Typically in a drive response system there is a generalized synchronization (GS) above a threshold value of the coupling when the response variables become implicit function of the drive \cite{rulkov}. When the largest conditional Lyapunov exponent (LCLE) $\lambda$ is positive, the dynamical states are desynchronized. The transition from asynchrony to bistable GS occurs gradually through a region where there are multiple coexisting attractors $C_{\mp, 0}$ as shown in Fig.~\ref{fig1}.

\begin{figure}[htb]
\centering
\includegraphics[width=.32\textwidth,angle=270]{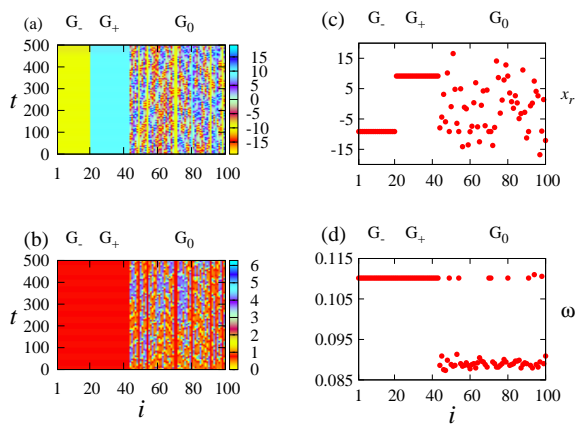}
\caption{Chimeric behaviour in an ensemble of $N = 100$ driven Lorentz oscillators at $\ep_1 = 0.70$. We plot (a) the time evolution of the $x_r$ variables and (c) their respective $x_r$ variables. We depict (b) the time evolution of phases and (d) their corresponding frequencies. Here $G_\mp$ represents two synchronized clusters and $(G_0)$ corresponds to desynchronized state.}
\label{fig2} 
\end{figure}

This analysis can be extended to study the dynamics of $N$ globally coupled Lorenz oscillators with linear diffusive coupling shown in Eq.~\ref{eq1}. The collective dynamics of Eq.~\ref{eq1} is a chimera as shown in Fig.~\ref{fig2} for $N = 100$ oscillators. 
The chimera state consists of three distinct sub-populations ($G_{\mp, 0}$) corresponding to the three distinct attractors $C_{\mp, 0}$ \cite{sangeeta-pre}. Fig.~\ref{fig2}(a) shows the time evolution of the $x_r$ variables where we observe that the entire population can be divided into three groups $(G_{\mp, 0})$. While in the first two groups $(G_\mp)$ the oscillators are synchronized, the oscillators in the third group ($G_0$) are desynchronized. This can also be seen by plotting the $x_r$ variables as shown in Fig.~\ref{fig2}(c). We also emphasize here that the oscillators in the first two groups ($G_\mp$) are in phase synchrony with each other. Evolution of phases of each oscillators is shown in Fig.\ref{fig2}(b). Their mean phase velocities (frequencies) given by $\omega_i = \frac {2\pi Q_i}{\Delta T}$ is plotted in Fig.~\ref{fig2}(d). Here $Q_i$ is the number of maxima of the time series $x_{r_i}(t)$ of the $i^ {th}$ oscillator in the time interval $\Delta T = 10^5$. Results are generated after removing initial transients for $2 \times 10^5$ times.

\section{ Effect of  Linear Augmentation}
\label{clinaug}

As argued earlier, one can destroy multistability in the system if the response is augmented to a linear system. This is possible because of the merging of the unstable fixed points \cite{sin}. The general form for a linearly augmented system is given by
\begin{eqnarray}
\label{eq3}
\dot{\overrightarrow{\bf{X}}}& = & F({\overrightarrow{\bf{X}}}) + \ep{\overrightarrow{\bf{U}}}\nonumber \\ 
\dot{\overrightarrow{\bf{U}}}& = & -K\overrightarrow{\bf{U}} - \ep{(\overrightarrow{\bf{X}} - B)}
\end{eqnarray}
$F(\overrightarrow{\bf{X}})$ represents the nonlinear system which is to be augmented, $\overrightarrow{\bf{X}} \in \mathbb{R}^m$,
$\varepsilon$ describes strength of feedback between the augmented oscillator and the linear system. $\dot{\overrightarrow{\bf{U}}} = -K\overrightarrow{\bf{U}}$ describes the dynamics of linear system, $K$ is the decay parameter \cite{resmi}, {$\overrightarrow{\bf{U}} = \left [u,0,0,0....\right]^T$}, $\ep{ (\overrightarrow{\bf{X}} - B)}$ provide sustained oscillations to the linear system. $B  = \left [b,0,0,0....\right] ^T$ is the control parameter of the augmented system which is close to the fixed points of the system \cite{sha, shr, sin} and $T$ indicates transpose.

\begin{figure}[htb]  
\centering
\scalebox{0.28}{\includegraphics{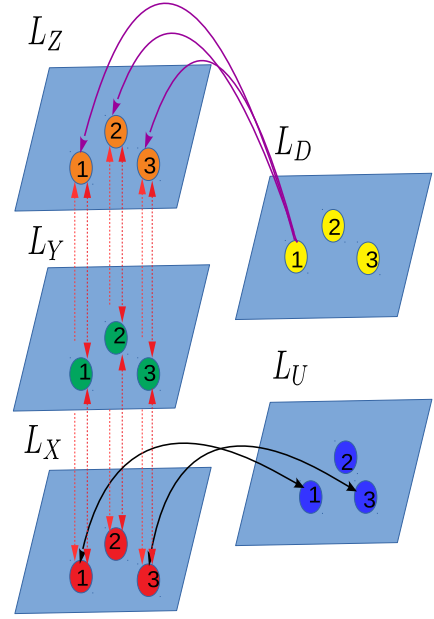}}
\caption{Extension of Fig.~\ref{fig0} after linear augmentation. Now, control layer $L_U$ is included with nodes of three linear systems. The bidirectional black color curves represent coupling between $X_r$ variables of desired number of response systems at layer $L_X$ and their corresponding linear systems at layer $L_U$ .}
\label{fig3}
\end{figure}

Here we target a given set of oscillators by adding another layer namely the control layer $L_U$ as shown in Fig.~\ref{fig3}. Layer $L_U$ contains a linear system on each node out of the $N$ linear systems. Note that in addition to the layers of the drive and response systems ($L_D, L_X, L_Y, L_Z$) as shown in Fig.~\ref{fig0}, we have added a control layer $L_U$ \ie now $l = m +2$. The dynamics of the drive-response system coupled to the control unit is given by (c.f. Eq.~\ref{eq1})
\begin{eqnarray}
\label{eq4}
\dot{x_d}&=& -{y_d} -{z_d}\nonumber \\ 
\dot{y_d}&=& {x_d}+a{y_d}\nonumber \\ 
\dot{z_d}&=&b+{z_d}({x_d}-c) \nonumber \\
\dot{x}_{r_i}&=& \sigma(y_{r_i} -x_{r_i} ) + \ep_2 E_{ij}  {u_j}\nonumber \\ 
\dot{y}_{r_i}&=&r x_{r_i}-y_{r_i}-x_{r_i}z_{r_i}\nonumber \\ 
\dot{z}_{r_i}&=&x_{r_i}y_{r_i}-\beta{z_{r_i}}+ \ep_1 A_{i1} \left(z_d - z_{r_i}\right)\nonumber\\  
\dot{u}_j&=&-Ku_j-\ep_2  E_{ij} {(x_{r_i} - B)}
\end{eqnarray}
Here, $i, j = 1, 2, 3 ,..., N$, $\ep_2$ is the coupling strength between the linear system and the response system, 
$B$ is close to the fixed points of the coupled drive response system and other parameters are same as described in Eq.~\ref{eq1}. $E_{ij}$ and $A_{i1}$ are $(N \times N)$ and $(N \times 1)$ adjacency matrices given by 
\[
\begin{bmatrix}
1 & 0  & 0 & \dots & 0 \\
0 & 0  & 0 & \dots & 0 \\
0 & 0  & 1 & \dots & 0 \\
. & .  & . & \dots & . \\
0 & 0 & 0  & \dots & 1
\end{bmatrix}
 \text{and}  
\begin{bmatrix}
1 &\dots & 1 
\end{bmatrix}^T
\] respectively. $E_{ij}$ represents the connection topology between the response layer $L_X$ and the control layer $L_U$. The dynamics at each node in the layer $L_U$ is given by the variable $u_j$. $E_{ij} = 1$, if there is a coupling between response system and its corresponding linear system, otherwise $E_{ij} = 0$. The topology for $A_{i1}$ matrix remains same as discussed earlier. The connection topology between the layers $L_{X}$ and $L_U$ is represented by  $A_{ij}^{[{L_X}{L_U}]}$. 
$A_{ij}^{[{L_X}{L_U}]}= E_{ij}^{[{L_X}{L_U}]} = E_{ij} = 1$ ($i = j$), if there is an interlayer connection between the nodes of $L_X$ and $L_U$ layers, otherwise $E_{ij} = 0$ \cite{fede,leyva, long, bocca, kivela}. In the absence of LA \ie $\ep_2 = 0$ ($A_{ij}^{[L_{X}L_{U}]} = E_{ij} = 0$), the dynamics of $N = 100$ coupled oscillators is a chimera state at $\ep_1 = 0.70$ 
as shown in Fig.~\ref{fig2}. In presence of LA, a given set of oscillators can be transformed to the desired state depending upon the value of $B$ at constant $K$ and $\ep_2$ and $E_{ij} = 1$ $(i = j)$ for oscillators of this set. For this purpose we determine fixed points of the coupled drive-response systems. Morphologically, the attractors $C_{\mp}$ are formed due to chaotic modulation of the fixed points \cite{fix} $\overrightarrow{\bf{X}}_{r_{i{_\mp}}}^*$ (determined below) in the response systems. Therefore, in this case $B$ is approximately equal to the fixed points of the coupled drive-response systems (Eq.~\ref{eq4}). Fixing $\ep_1 = 0.70$ for which chimera states have been obtained 
(Fig.~\ref{fig2}) and keeping decay parameter $K = 5$,  desired number of response systems are connected to the linear systems.

At equilibrium, the coupled response systems can be written as
\begin{eqnarray}
\label{eq5}
 \sigma(y_{r_i}^* -x_{r_i}^*) = 0 \nonumber\\
 r x_{r_i}^*-y_{r_i}^*-x_{r_i}^*{z_{r_i}^*} = 0 \nonumber\\
 x_{r_i}^*y_{r_i}^*-\beta z_{r_i}^* +\varepsilon_1 A_{i1}(z_d-z_{r_i}^*) = 0
\end{eqnarray}
After solving we get three sets of equilibrium points:
$\overrightarrow{\bf{X}}_{r_{i_0}}^*=(x_{r_{i_0}}^*, y_{r_{i_0}}^*, z_{r_{i_0}}^*)=(0, 0, 0)$ situated at the origin and another two sets of equilibrium points are $\overrightarrow{\bf{X}}_{r_{i_\mp}}^*=(x_{r_{i_\mp}}^*,y_{r_{i_\mp}}^*, z_{r_{i_\mp}}^*)=
(\mp\sqrt{(\beta+\ep_1 A_{i1})(r-1)-\ep_1 A_{i1}z_d},r-1)$, where $x_{r_{i_\mp}}^*=y_{r_{i_\mp}}^*$. We make the following choices for $B$: $B \simeq B_\mp= x_{r_{i_\mp}}^*=\mp\sqrt{(\beta + \ep_1 A_{i1})(r-1)-\ep_1 A_{i1}z_d}$ and $B = B_0$, where $x_{r_{i_-}}^* << B_0 << x_{r_{i_+}}^*$. Therefore, for $B \simeq B_\mp$, the system dynamics moves to the attractor $C_\mp$ (Figs.~\ref{fig1}(a), (b)) and for $B = B_0$, the system dynamics evolves towards the $C_0$ attractor (Fig.~\ref{fig1}(c)). Note that for $B = B_\mp$, the system dynamics is a fixed point $x^*_{r_{i _\mp}}$. Therefore, we consider $B$ to be approximately equal to $B_\mp$ \ie $B\simeq B_\mp$.

\begin{figure}[htb]  
\centering
\scalebox{0.45}{\includegraphics{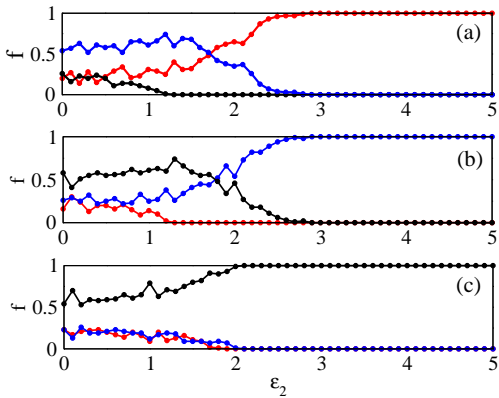}}
\caption{Fraction of initial conditions going to $C_{\mp,0}$ attractors represented by red (light gray), blue (dark gray) and black curves, respectively at decay parameter $K = 5$ and at coupling strength $\ep_1 = 0.70$. Control parameters are set at (a) $B \simeq B_-$, (b) $B \simeq B_+$ and (c) $B = B_0$. To determine these fractions of initial conditions we coupled single linear system to a Lorentz system driven by a R\"ossler system with $N = 1$ (Eq.~\ref{eq4}) over $500$ different initial conditions.} 
\label{fig4}
\end{figure}
Fig.~\ref{fig4} represents fraction of initial conditions going to the attractors $C_{\mp, 0}$ represented by red (light gray), blue (dark gray) and black curves, respectively at $\ep_1 = 0.70$ and $K = 5$. Control parameters in Figs.~\ref{fig4} (a), (b), (c) are $B \simeq B_ {\mp, 0}$, respectively. These results correspond to the case when there is only one $(N = M = 1)$ linearly augmented Lorenz system driven by  R\"ossler drive which is given by equation Eq.~\ref{eq4}. In Fig.~\ref{fig4}, for the value of coupling strength $\ep_2\gtrsim 3$, all initial conditions move to a desired state depending upon the value of $B$. Therefore, the region $\ep_2\gtrsim 3$ is suitable for controlling the dynamics. Thus, in an ensemble of oscillators, we can use this scheme to obtain the desired collective state. 

Setting the augmentation strength at $\ep_2 = 4$ following cases have been explored to study the collective dynamics of Eq.~\ref{eq4} for $N = 100$ oscillators.

\subsection{Case I}
Here $M = 50$ nodes in the $L_X$ layer are augmented to $M = 50$ nodes of the $L_U$ layer while remaining $n = N - M = 50$ oscillators remain unaugmented.
\begin{figure}[htb]
\centering
\includegraphics[width=.32\textwidth,angle=270]{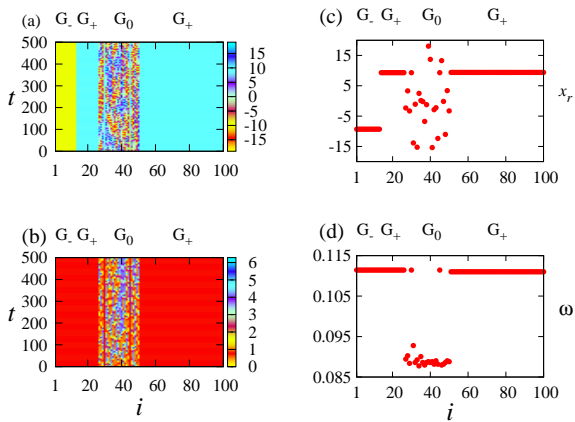}
\caption{Plots similar to Fig.~\ref{fig2}, showing the controlled chimera states for $N = 100$ oscillators. We represent (a) space-time plot and (c) their corresponding plot for $x_r$ variables. We describe (b) time evolution of phases and (d) their corresponding frequencies. All the results are obtained at $\ep_1 = 0.70$, $K = 5$, $B \simeq B_+$ and $\ep_2 = 4$. There are three synchronized clusters $(G_{\mp}, G_+)$ and  a desynchronized state $(G_0)$. Here, first $n = 50$ unaugmented oscillators exhibits chimera state and remaining $M = 50$ augmented oscillators are forced to go to the group of $G_+$ oscillators where they become synchronized.}
\label{fig5} 
\end{figure}
Setting $B\simeq B_+$, we observe that the resulting dynamics of oscillators can be given on the attractor $C_+$ and this set forms the group $G_+$. Fig.~\ref{fig5}(a) shows time evolution of the $x_r$ variables for $N = 100$ oscillators where $n = 50$ unaugmented oscillators remain in chimera state. The remaining set of $M = 50$ oscillators exhibit dynamics that is identical to that of synchronized $(G_+)$ cluster where $C_+$ attractors are stabilized. Fig.~\ref{fig5}(c) describes their corresponding plot for the $x_r$ variables at a given time. In Fig.~\ref{fig5}(b) we plot phase-time relationship with their average phase velocities (frequencies) plotted in  Fig.~\ref{fig5}(d). This again shows that the oscillators in the groups $G_+$ and $G_-$ are in phase synchrony. Similarly these selected set of oscillators can be linear augmented to follow the dynamics of the oscillators present in the $G_-$ (synchronized) or $G_0$ (desynchronized) groups by setting $B \simeq B_-$ or $B = B_0$.

\subsection{Case II} 
In this case we couple each node of the $L_X$ layer to a node in the $L_U$ layer. We observe that the entire chimera state present in the network of response systems can be destroyed.
\begin{figure}[htb]
%\label{figaa}
\centering
\includegraphics[width=.32\textwidth,angle=270]{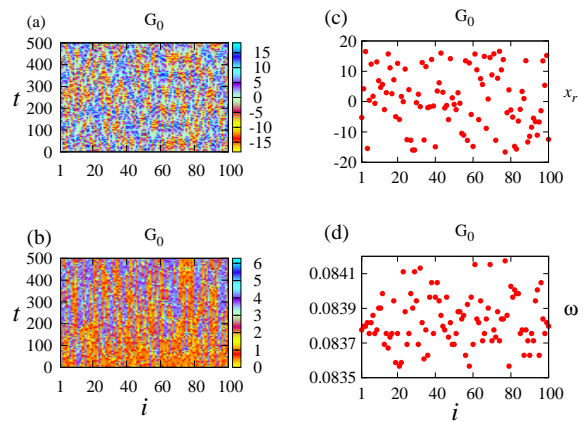}
\caption{All the $N = 100$ oscillators are forced to follow the dynamics of $C_0$ attractor where they become desynchronized  at the parameter values $\ep_1 = 0.70$, $K = 5$, $B = B_0$ and $\ep_2 = 4$. We represent (a) the time evolution of the $x_r$ variables and (b) their corresponding plot for $x_r$ variables. We plot (b) the time evolution of phases and (d) their corresponding frequencies.}
\label{fig6} 
\end{figure}
In Fig.~\ref{fig6} using $B = B_0$, we augment $M = N = 100$ oscillators so that all the oscillators go to the desynchronized ($G_0$) state and follow the dynamics of $C_0$ attractor. Fig.~\ref{fig6}(a) represents space-time plot and Fig.~\ref{fig6}(c) indicates their corresponding plot for $x_r$ variables for $N = 100$ oscillators. Similarly, in Fig.~\ref{fig6}(b) we show phase-time plot and Fig.~\ref{fig6}(d) represents their
corresponding frequencies. Using similar arguments one may also make the transition to a completely synchronized state by setting $B \simeq B_-$ or $B \simeq B_+$ in which case all the oscillators would either form $G_-$ or $G_+$ group. Thus, one can transform the whole dynamics present in the chimera state to a single dynamics thereby annihilating chimera states.

\section{Selection of control region}
\label{control}

\begin{figure}[htb]
\centering
    \includegraphics[width=.32\textwidth,angle=270]{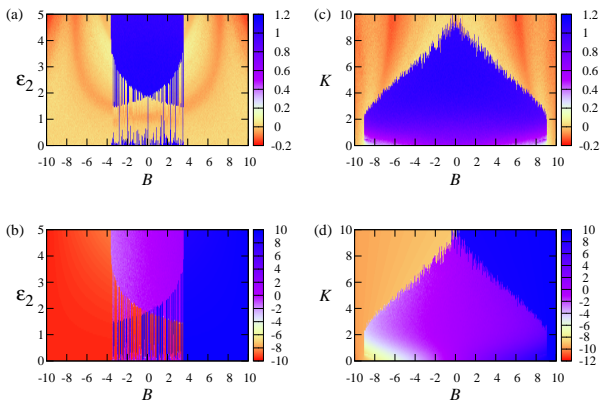}
  \caption{ Variation of (a) largest conditional Lyapunov exponent and (b) average value of $x_r$ variable over time $t$ $({<x_r>}_t)$ in parameter space $B-\ep_2$ at $K = 5$. Variation of (c) largest conditional Lyapunov exponent and (d) average value of $x_r$ variable over time $t$ $({<x_r>}_t)$ in parameter space $K-B$ at $\ep_2 = 4$ for the response system in Eq.~\ref{eq4} with $N = 1$ at $\ep_1 = 0.70$}
\label{fig7} 
\end{figure} 
We plot the largest conditional Lyapunov exponent (LCLE) \cite{carrol,koca} and average value of $x_r$ variable ${<x_r>}_t$ in the $B-\ep_2$ space in Figs.~\ref{fig7}(a) and (b) respectively, at $K = 5, \ep_1 = 0.7$ and $M  = N = 1$. For $\ep_2 < 3$ and $-3 < B <3$, we observe a mixed region where the LCLE exhibits both negative and positive values implying that the initial conditions may evolve to any of the coexisting attractors $C_{\mp,0}$. Thus, effective control can be achieved only for the $\ep_2 > 3$. This observation is also consistent with the values of ${<x_r>}_t$ plotted in Fig.~\ref{fig7}(b). In the region $B < -3$ and $B > 3$, dark orange (dark gray) shaded region corresponds to the negative values of LCLE  and the initial conditions asymptote only to the attractors $C_\mp$ that remains synchronized. Other light orange (light gray) shaded region corresponds to the positive LCLE implying desynchronized dynamics. In Figs.~\ref{fig7}(c) and (d), we plot LCLE and ${<x_r>}_t$ respectively in the $B-K$ parameter space. The cone shaped region in Fig.~\ref{fig7}(c) represents the positive values of LCLE, where the dynamics evolves to the $C_0$ attractor (Fig.~\ref{fig7}(d)). Therefore, the dynamics in this region remains desynchronized. Outside this cone for dark orange (dark gray) shaded region, the LCLE becomes negative and the resulting attractors are either $C_-$ or $C_+$.

\begin{figure}[htb]
\centering
    \includegraphics[width=.32\textwidth,angle=270]{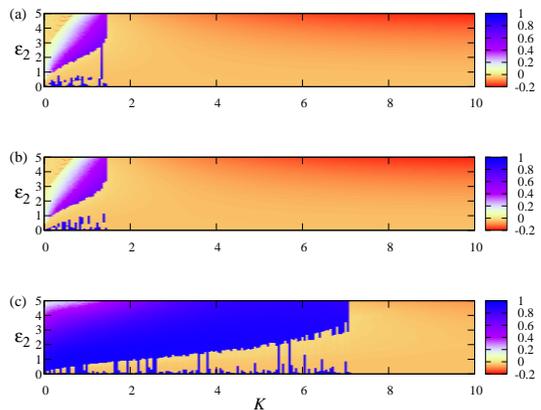}
  \caption{Variation of largest conditional Lyapunov exponent in $K-\ep_2$ parameter space at (a) $B \simeq B_-$, (b) $B \simeq B_+$ and (c) $B = B_0$, respectively. These are plotted using equation Eq.~\ref{eq4} with $N = 1$ at $\ep_1 = 0.70$. Side bar represents largest conditional Lyapunov exponent of the response system.}
\label{fig8} 
\end{figure} 
We also explore $K-\ep_2$ parameter space by plotting the LCLE for $B \simeq B_{\mp, 0}$ in Figs.~\ref{fig8}(a), (b) and (c) respectively. 
For $B \simeq B_{\mp}$ (Figs.~\ref{fig8} (a) and (b)) the LCLE always remains negative for $K > 1.5$. This suggests that the dynamics is always synchronized and corresponds to the attractors $C_\mp$. In the region $K < 1.5$, LCLE may be positive or negative and therefore initial conditions may evolve to any of the three coexisting attractors $C_{\mp, 0}$. This is the region where chimera states are observed. In Fig.~\ref{fig8}(c) for $K < 7$, except the orange (light gray) shaded region one can observe positive value of LCLE for which initial conditions evolve towards $C_0$ attractor at $B = B_0$. Thus one may infer that the dynamics cannot be controlled in the region $K < 1.5$ as shown in Figs.~\ref{fig8}(a), (b) and for orange (light gray) shaded region as shown in Fig.~\ref{fig8}(c).

\section{Robustness and stability of different states before and after Linear Augmentation}
\label{robust}

To verify the robustness of chimera states before LA (Eq.~\ref{eq1} with $N = 2$) and after LA (Eq.~\ref{eq4} with $N = 2$), we explore the changes in the basins of attraction of coupled drive-response systems for different values of $\ep_2$ and at a constant value of control parameter $B$. In this case we choose $B = B_0$ and observe the changes in the basins of attraction.
\begin{figure}[htp]  
\centering
 \begin{tabular}{@{}c@{}}
\scalebox{0.45}{\includegraphics{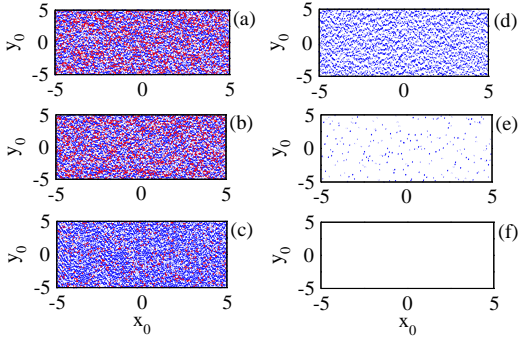}}
 \end{tabular}
\caption{Basins of attraction for coupled drive-response oscillators with $N = 2$ (a) before LA using Eq.~\ref{eq1} and (b)-(f) after LA using Eq.~\ref{eq4}. These are plotted for augmentation strength (a) $\ep_2 = 0.0$, (b) $\ep_2 = 0.1$, (c) $\ep_2 = 0.5$, (d) $\ep_2 = 2.0$, (e) $\ep_2 = 3.0$ and (f) $\ep_2 = 4.0$. The other parameters are fixed at $\ep_1 = 0.7$, $B = B_0$, $K = 5$. Here, red (light gray) and blue (dark gray) regions represent the initial conditions that evolve towards the attractors $C_\mp$. White region corresponds to the initial conditions evolving towards the attractor $C_0$.}
\label{fig9}
\end{figure}
As shown in Fig.~\ref{fig9}(a), the basins of attraction for coupled drive-response systems (Eq.~\ref{eq1} with $N = 2$) is riddled and completely interwoven in a complex manner at $\ep_1 = 0.70$ and $\ep_2 = 0$. From Fig.~\ref{fig9}(a), we may conclude that two randomly selected nearby initial conditions may asymptote to different regimes that may be synchronized or desynchronized or we may say that chimera states may occur for random \cite{maistrenko, ujjwal} or quasirandom \cite{showalter} initial conditions. It is very likely that the basins will be even more intertwined for large volumes and larger $N$ values. It has been shown that the basin structure of different attractors in coupled systems is complex \cite{camargo, ujjwal}. Here, red (light gray), blue (dark gray) and white regions represent initial conditions which lead dynamics to the $C_{\mp, 0}$ attractors, respectively. Figs.\ref{fig9}(b-f) represents the basins of attraction for the coupled systems after LA (Eq.~\ref{eq4} with $N = 2$) for different values of augmentation strengths ($\ep_2$). In this case we observe that with increasing $\ep_2$ the initial conditions corresponding to attractor $C_0$ increases because the control parameter of the augmented system is set at $B = B_0$.

We calculate the fraction of initial conditions $f_{ic}$ as a function of transversal distance $ S_{td} = \mid z_{r_1} - z_{r_2} \mid$ \cite{sangeeta-chaos, camargo} from the synchronization manifold using Eq.~\ref{eq4} with $N = 2$ and $B = B_0$ for different values of coupling strength $\ep_2$. For each value of $S_{td}$, we consider an ensemble of $10^3$ initial conditions drawn from the riddled region with $x_d = y_d = z_d = 1$, $x_{r_1} = y_{r_1} = 1$, $x_{r_2} = y_{r_2} = 1$ and $z_{r_1}, z_{r_2}$ are chosen from the interval $[7,9]$ at $\ep_1 = 0.7$ and calculate the fractions leading to different attractors $(C_{\mp, 0})$. 
\begin{figure}[htb]  
\centering
\scalebox{0.40}{\includegraphics{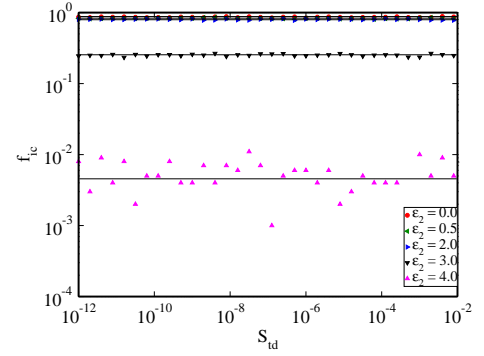}}
\caption{Variation of the fraction of initial conditions $f_{ic}$ going to different attractors $(C_{\mp, 0})$ as a function of transversal distance $ S_{td} = \mid {z_{r_1} - z_{r_2}}\mid$. Using Eq.~\ref{eq4} with $N = 2$ for each value of $ S_{td} $, we consider $10^3$ initial conditions with $x_d = y_d = z_d = 1$, $x_{r_1} = y_{r_1} = 1$, $x_{r_2} = y_{r_2} = 1$ and $z_{r_1}, z_{r_2}$ are chosen from the interval $[7,9]$ at $\ep_1 = 0.7$.}
\label{fig123}
\end{figure}
In Fig.\ref{fig123}, we change the value of augmentation strength $\ep_2$ for a fixed value of coupling strength $\ep_1$ and $B = B_0$. In the absence of LA \ie with $\ep_2 = 0.0$, the value of $f_{ic}$ does not change with $S_{td}$ and there is always a constant value $f_{ic} \simeq 1$. However, as one increases the value of $\ep_2$, we observe that $f_{ic}$ decreases indicating that lesser number of initial conditions go to different attractors. This is because of the fact that by setting $B = B_0$, the initial conditions corresponding to $C_0$ attractor increases with increase in $\ep_2$. This observation is found to be consistent with Fig.~\ref{fig4}(c) and Fig.~\ref{fig9}. Thus, it can be inferred from Fig.~\ref{fig123} that with increasing strength of LA, there is a uniform decrease in the initial conditions going to different attractors $(C_{\mp, 0})$.

We also compute largest transversal LE by considering the time evolution of the infinitesimal displacement along a direction transversal to the synchronization manifold \cite{pecora-chaos, sangeeta-chaos}. This was calculated by writing Eq.~\ref{eq4} for $i = j = 2$ given by.
\begin{eqnarray}
\label{eqtr}
\dot{x_d}&=& -{y_d} -{z_d}\nonumber \\ 
\dot{y_d}&=& {x_d}+a{y_d}\nonumber \\ 
\dot{z_d}&=&b+{z_d}({x_d}-c) \nonumber \\
\dot{x}_{r_1}&=& \sigma(y_{r_1} -x_{r_1} ) + \ep_2 E_{11}{u_1}\nonumber \\ 
\dot{y}_{r_1}&=&r x_{r_1}-y_{r_1}-x_{r_1}z_{r_1}\nonumber \\ 
\dot{z}_{r_1}&=&x_{r_1}y_{r_1}-\beta{z_{r_1}}+ \ep_1 A_{11} \left(z_d - z_{r_1}\right)\nonumber\\  
\dot{u}_1&=&-Ku_1-\ep_2 E_{11}{(x_{r_1} - B)}\nonumber \\
\dot{x}_{r_2}&=& \sigma(y_{r_2} -x_{r_2} ) + \ep_2 E_{22}{u_2}\nonumber \\ 
\dot{y}_{r_2}&=&r x_{r_2}-y_{r_2}-x_{r_2}z_{r_2}\nonumber \\ 
\dot{z}_{r_2}&=&x_{r_2}y_{r_2}-\beta{z_{r_2}}+ \ep_1 A_{21}\left(z_d - z_{r_2}\right)\nonumber\\  
\dot{u}_2&=&-Ku_2-\ep_2 E_{22}{(x_{r_2} - B)}
\end{eqnarray} 
where $A_{11} = A_{21} = E_{11} = E_{22} = 1$ and other parameters are same as discussed earlier. If we perform the changes in variables 
 \beqr
 x=x_{r_2}-x_{r_1};y=y_{r_2}-y_{r_1};z=z_{r_2}-z_{r_1};u=u_2-u_1 \nonumber \\ 
 X=x_{r_1}+x_{r_1};Y=y_{r_2}+y_{r_1};Z=z_{r_2}+z_{r_1};U=u_2+u_1 
 \eqnr
after which the coupled equations (Eq.~\ref{eqtr}) are given by 
 \beqr
 \label{eqnl}
 \dot{x}&=& \sigma(y -x); ~~~~~\dot{y}= r x- y-(Xz+Zx) \nonumber \\
\dot{z}&=& (\beta +2\ep_1)z +Xy+Yx; ~~~~~~ \dot{u}=-Ku - 2\ep_2 x \nonumber \\
 \dot{X}&=& \sigma(Y -X); ~~~~~\dot{Y}= r X- Y-(XZ+xz) \nonumber \\
\dot{z}&=& -\beta Z +XY+xy;~~\dot{U}=-KU - \ep_2(X-2B)
\eqnr

In these transformed systems we have a new set of coordinates in which three coordinates are on the synchronization manifold $(X, Y, Z)$ and the other 
three are on the transverse manifold $(x, y, z)$. We now calculate the largest transverse Lyapunov exponent (TLE) which describes whether perturbations transverse to the synchronization manifold grow or shrink as $t\rightarrow \infty$. In order to determine TLE we put $x = y = z = 0$; $X = Y = 1$ and Z is chosen randomly.

\begin{figure}[htb]  
\centering
\scalebox{0.55}{\includegraphics{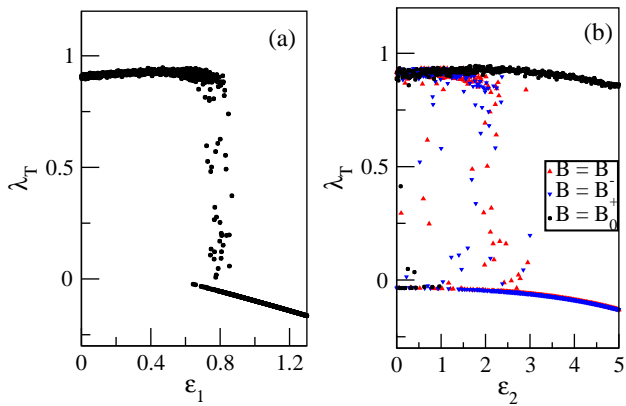}}
\caption{Variation of the largest transversal Lyapunov exponent $(\lambda_T)$ with respect to coupling strength $\ep_1$ corresponding to Eq.~\ref{eq1} for $N = 2$ at $\ep_2 = 0$ i.e. before LA and (b) Variation of the largest $\lambda_T$ corresponding to Eq.~\ref{eq4} with $N = 2$, $\ep_1 = 0.70$ at different values of $\ep_2$ after LA. (b) Red (light gray), blue (dark gray) and black curves represent $(\lambda_T)$ for $B \simeq B_{\mp, 0}$, respectively at $K = 5$. These figures are plotted for 30 different initial conditions.}
\label{fig14}
\end{figure}
In Fig.~\ref{fig14}(a) we plot the variation of largest TLE with the coupling strength in the absence of linear augmentation ($\ep_2 = 0$). For smaller values of $\ep_1$, largest TLE is positive indicating desynchronized states. For intermediate values of $\ep_1$, TLE is both positive and negative for different initial conditions. This indicates the coexistence of coherent and incoherent dynamics for different initial conditions. If we increase $\ep_1$ further we observe that largest TLE is negative indicating that the dynamics is synchronized. To see the effect of linear augmentation, we fix $\ep_1 = 0.7$ and plot largest TLE w.r.t the augmentation strength $\ep_2$ in Fig.~\ref{fig14}(b). We observe that there are three curves (red (light gray), blue (dark gray), black) corresponding to the three values of the control parameter $B \simeq B_{\mp, 0}$. For $\ep_2 \lesssim 3$, we can see a mixed region with both positive and negative largest TLE, irrespective of the value of $B$. Therefore, this region is not preferable to achieve the desired dynamics. At $\ep_2 > 3$, largest TLE corresponding to $B = B_0$ is always positive (black curve) implying that the resulting dynamics is always desynchronized and for the control parameter $B \simeq B_\mp$, largest TLE becomes negative (red (light gray), blue (dark gray) curves) indicating synchronized dynamics. This is the region where one can force the dynamics to a desired states for proper choice of $B$.  

We also explore the idea of the master stability function (MSF) \cite{pecora-msf, huang} to study the effect of linear augmentation in the multilayer network. One can write a network of $N$ coupled oscillators as $\frac{d {\bf X}_i}{dt} = {\bf F}({\bf X}_i)-\ep \sum_{j=1}^N G_{ij}{\bf H}({\bf X}_j)$, where $\ep$ is a global 
coupling parameter, ${\bf H(X)}$ is a coupling function and $\bf {G}$ represents a coupling matrix determined by the connection topology. The variational 
equations $\frac{d\delta{\bf X}_i}{dt} = {\bf DF(s)}\cdot\delta {\bf X}_i-\ep 
\sum_{j=1}^{N}G_{ij}{\bf DH(s)}\cdot\delta
{\bf X}_j$ govern the time evolution of the set of infinitesimal vectors about the synchronous solution. The generic form of all decoupled blocks is given by 
 \begin{eqnarray}
 \label{eq7}
 \frac{d\delta \bf{y}}{dt}=\left [\bf {DF(s)}-\kappa \bf {DH(s)} \right]\cdot\delta \bf {y}.
 \label{eqmsf}
 \end{eqnarray}
where, $\kappa$ is a normalized coupling parameter. The largest LE for this equation $\lambda_M(\kappa)$ gives the MSF which describes the linear stability of the synchronized dynamics. 

\begin{figure}[htb]  
\centering
 \begin{tabular}{@{}c@{}}
\scalebox{0.55}{\includegraphics{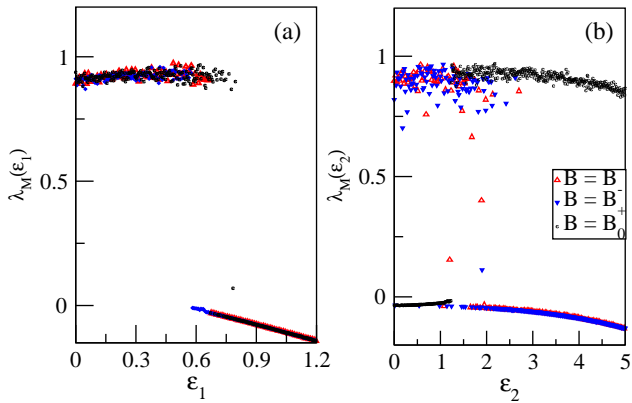}}
 \end{tabular}
\caption{(a) Variation of the MSF ($\lambda_M(\ep_1)$) corresponding to Eq.~\ref{eq1} with $\ep_1$ at $\ep_2 = 0$ i.e. before linear augmentation and (b) variation of the MSF corresponding to Eq.~\ref{eq4} with $\ep_2$ at $\ep_1 = 0.70$ after LA for 100 different initial conditions. (a) Red (light gray), blue (dark gray) and black curves represent $\lambda_M(\ep_1)$ for $C_{\mp, 0}$ attractors, respectively. (b) Red (light gray), blue (dark gray) and black curves represent $\lambda_M(\ep_2)$ for $B \simeq B_{\mp, 0}$, respectively at $K = 5$.}
\label{fig16}
\end{figure}

In order to visualize the stability of different states present in the multistable network it is important that one should rather calculate the MSF of individual attractors \cite{bocaletti-msf}. For coupled drive-response flows before LA (Eq.~\ref{eq1}), we calculate $\lambda_M(\ep_1)$ in Fig.\ref{fig16}(a). For smaller values of $\ep_1$ one may observe that the MSF always remains positive (red (light gray), blue (dark gray), black curves) for all the three attractors $C_{\mp, 0}$ indicating unstable desynchronized states. As the value of $\ep_1$ is increased the MSF becomes negative in the region $\ep_1 \simeq 0.60 - 0.80$. Interestingly, it can be seen that in this region $\lambda_M(\ep_1)$ has different behaviour for different attractors, this implies that for a given value of $\ep_1$, the dynamics over a given attractor may be synchronized, while for other attractors it may not be synchronized. This results in a state where we have a coexisting coherent and incoherent regions. As one increases the value of $\ep_1$ further, the MSF for all attractors becomes negative implying stable synchrony.
After LA the system dynamics depends on control parameter $B$ of the linear system in layer $L_U$ and becomes completely independent of the initial conditions. Therefore, after LA, we calculate MSF, $\lambda_M(\ep_2)$ in Fig.~\ref{fig16}(b) corresponding to different values of $B$ for $K = 5$ and $\ep_1 = 0.70$. Similar to Fig.~\ref{fig14}(b), here also for $\ep_2 \lesssim 3$, we can observe a mixed states of both positive and negative values of $\lambda_M(\ep_2)$ leading the system dynamics to the stable and unstable states irrespective of the value of $B$. As one increases $\ep_2 > 3$, value of $\lambda_M(\ep_2)$ corresponding to $B \simeq B_{\mp}$ (red (light gray), blue (dark gray) curves) becomes negative indicating stable synchronized dynamics and $\lambda_M(\ep_2)$ is positive for $B = B_0$ (black curve) leading to desynchronized dynamics of the network. Therefore, in order to control the chimera or in other words to move the dynamics to a single state one must avoid the region for $\ep_2 \lesssim 3$.
 
\section{Conclusion}
\label{conclusion}
In this work we have examined the emergence of dynamical chimeras in the frame work of multilayer networks. These chimera states can further be controlled by introducing a control layer where each node is a linear system. If we  selectively switch on the couplings between these nodes then one can effectively control the resulting collective dynamics. Since the control in this case takes place at the level of individual nodes, it is possible to control the dynamics and hence control the spatial locations in the chimera states. Our results suggest that by fixing decay parameter $K$  and coupling
strengths ($\ep_1$ and $\ep_2$), it is possible to shift the desired number of  oscillators to a particular state depending upon the value of $B$. We also observe that after LA a particular (entire) set of the population of oscillators settles into a single state (synchronized or desynchronized) and becomes totally independent of the coupling mechanisms and initial conditions. By considering the Lorenz dynamics in the three layers driven by the  R\"ossler oscillator, we observe that the basins are riddled. In presence of connections with the control layer, riddling is reduced. The riddled nature of the basins of attraction is verified by showing that the fraction of initial conditions going to different attractors obey power law. For very large augmentation strengths, the basins were found to be dominated by initial conditions going to the desired attractor.

We have also calculated transverse Lyapunov exponents which is negative for synchronized motion while it is positive if the dynamics is desynchronized. In presence of LA we observe that the largest TLE reduces to three curves for different values of the control parameter. The resulting curve suggests synchrony for the negative values of TLE and desynchrony otherwise. By calculating Master Stability Function, we confirmed the stability of synchronized state. Since the MSF in presence of LA has different values for different attractors we conclude that the dynamics of the synchronized state is stable for which the MSF is negative.

The results presented here are general in the sense that they can be applied to an ensemble of oscillators where chimera states are observed through multistability. Throughout this study, we have taken $x_r$ variable for linear augmentation. We would like to emphasize here that one may choose other variables as well. For generality we have checked our results for different drives in presence of mutually coupled global and nonlocal couplings. In each of these cases we observed that our results apply quite generally. 

The techniques outlined in this work will in general be helpful in engineering collective dynamics in multilayer networks where some layers are not
accessible. In particular, LA may be helpful in controlling the spatial locations in the chimera states. There are instances, namely in power 
grid and neuronal systems where it is difficult to access internal parameters of the system. In such cases one may employ these techniques to ensure
effective control.

\section*{Acknowledgment:} AAK acknowledges UGC, India for the financial support. HHJ would like to thank the UGC, India for the award of grant no. F:30-90/2015 (BSR). We also thank Ram Ramaswamy for useful discussions.


\begin{thebibliography}{abc99}
\bibitem{kuramoto} Y. Kuramoto and D. Battogtokh, Nonlinear Phenom. Complex Syst. {\bf 5}, 380 (2002).%1
\bibitem{strogatz} D. M. Abrams and S. H. Strogatz, \prl{93}{174102}{2004}.%2
\bibitem{thutupalli} E. A. Martens, S. Thutupalli, A. Fourri\'ere, and O. Hallatschek, \pnas{110}{10563}{2013}.%3
\bibitem{kapitaniak} T. Kapitaniak, P. Kuzma, J. Wojewoda, K. Czolczynski, and  Y. Maistrenko, Sci. Rep. {\bf 4}, {6379} (2014).
\bibitem{battagila}D. Battagila, N. Brunel, and D. Hansel, \prl{99}{238106}{2007}.%5
\bibitem{vicente} R. Vicente, L. L. Gollo, C. R. Mirasso, I. Fischer, and P. Gorden, \pnas{105}{17157}{2008}.%6
\bibitem{rattenborg} N. C. Rattenborg, C. J. Amlaner, and S. L. Lima, Neurosci. Biobehav. Rev. {\bf 24}, 817 (2000).%7
\bibitem{laing} C. R. Laing and C. C. Chow, Neural Comput. {\bf 13}, 1473 (2001); 
C. R. Laing, W. C. Troy, B. Gutkin, and G. B. Ermentrout, SIAM J. Appl. Math. {\bf 63}, 62 (2002).%8
\bibitem{wiesenfield} K. Wiesenfield, P. Colet, and S. H. Strogatz, \pre {76}{404}{1996}.%9
\bibitem{mazouz} N. Mazouz, G. Fl\"atgen, and K. Krischer, \pre{55}{2260}{1997}.%10
\bibitem{garcia} V. Garcia-Morales and K. Krischer, \prl{100}{054101}{2008}.%11
\bibitem{chandrasekar} V. K. Chandrasekar, R. Gopal, A. Venkatesan, and M. Lakshmanan, \pre{90}{062913}{2014}.%12
\bibitem{sangeeta-pre} S. R. Ujjwal, N. Punetha, A. Prasad, and R. Ramaswamy, \pre{95}{032203}{2017}.%13
\bibitem{anjuman} A. A. Khatun and H. H. Jafri, arXiv:1903.02739 .%14
\bibitem{yclai}Y. C. Lai, C. Grebogi, J. A. Yorke, and S. C. Venkataramani, \prl{77}{55}{1996}.%15
\bibitem{viana} S. Camargo, R. L. Viana, and C. Anteneodo, \pre{85}{036207}{2012}.%16
\bibitem{sangeeta-chaos} S. R. Ujjwal, N. Punetha, R. Ramaswamy, M. Agarwal, and A. Prasad, \chaos{26}{06311}{2016} and references therein.%17
\bibitem{bick} C. Bick, C. Kolodziejski, and M. Timme, \chaos{24}{033138}{2014}.%18
\bibitem{sieber}J. Sieber, O. E. Omel\'chenko, and M. Wolfrum, \prl{112}{054102}{2014}.%19
\bibitem{omel16}I. Omelchenko, O. E. Omel\'chenko, A. Zakharova, M. Wolfrum, and E. Sch\"oll, \prl{116}{114101}{2016}.%20
\bibitem{omel18}I. Omelchenko, O. E. Omel\'chenko, A. Zakharova, and E. Sch\"oll, \pre{97}{012216}{2018}.%21
\bibitem{omel19}I. Omelchenko, T. H\"ulser, A. Zakharova, and E. Sch\"oll, Front. Appl. Math. Stat. {\bf 4}, 67 (2019).%22
\bibitem{gjurchinovski} A. Gjurchinovski, E. Sch\"oll, and A. Zakharova, \pre{95}{042218}{2017}.%23
\bibitem{bera-control}B. K. Bera, D. Ghosh, P. Parmananda, G. V. Osipov, and S. K. Dana, \chaos{27}{073108}{2017}.%24
\bibitem{sha} P. R. Sharma, A. Sharma, M. D. Shrimali, and A. Prasad, \pre{83}{067201}{2011}.%25  
\bibitem{shr}P. R. Sharma, M. D. Shrimali, A. Prasad, and U. Feudel, \pla{377}{2329}{2013}.%26
\bibitem{sin} P. R. Sharma, A. Singh, A. Prasad, and M. D. Shrimali, \epj{223}{1531}{2014}.%27
\bibitem{pra}P. R. Sharma, M. D. Shrimali, A. Prasad, N. V. Kuznetsov, and G. A. Lenov, \ijbc{25(4)}{1550061}{2015}.%28
\bibitem{sevil} R. Sevilla-Escoboza, R. Guti\'errez, G. Huerta-Cuellar, S. Boccaletti, J. G\'omez-Garde\~nes, A. Arenas, and J. M. Buld\'u, 
\pre{92}{032804}{2015}.%29
\bibitem{bocca} S. Boccaletti, G. Bianconi, R. Criado, C. I. del Genio, J. G\'omez-Garde\~nes, M. Romance, I. Sendi\~na-Nadal,
Z. Wang, and M. Zanin, Physics Reports {\bf{544}}, 1-122 (2014).%30
\bibitem{kivela} M. Kivel\"a, A. Arenas, M. Barthelemy, J. P. Gleeson, Y. Moreno, and M. A. Porter, Journal of Complex Networks 
{\bf{2}}, 203-271, (2014).%31
\bibitem{pecora-msf}L. M. Pecora and T. L. Carroll, \prl{80}{2109}{1998}.%32
\bibitem{huang}L. Huang, Q. Chen, Y. C. Lai, and L. M. Pecora, \pre{80}{036204}{2009}.%33
\bibitem{bocaletti-msf}R. Sevilla-Escoboza, J. M. Buld\'u, A. N. Pisarchik, S. Boccaletti, and R. Guti\'errez, \pre{91}{032902}{2015}.%34
\bibitem{gamb}L. V. Gambuzza, M. Frasca, and J. G\'omez-Garde\~nes, Euro. Phys. Lett. {\bf{110}}, 20010 (2015).%35
\bibitem{leyva} I. Leyva, R. Sevilla-Escoboza, I. Sendiña-Nadal, R. Guti\'errez, J.M. Buld\'u, and S. Boccaletti, 
Scientific Reports {\bf{7}}, 45475 (2017).%36
\bibitem{long} L. Tang, X. Wu, J. L\"u, J. Lu, and R. M. D'Souza, \pre{99}{012304}{2019}.%37
\bibitem{fede} F. Battiston, V. Nicosia, and V. Latora, \pre{89}{032804}{2014}.%38
\bibitem{rulkov} N. F. Rulkov, M. M. Sushchik, L. S. Tsimring, and H. D. Abarbanel, \pre{51}{980}{1995}.%39
\bibitem{resmi} V. Resmi, G. Ambika, and R. E. Amritkar, \pre{81}{046216}{2010}.%40
\bibitem{fix}C. Spparow, {\it The Lorentz equations: Bifurcations, chaos and strange attractors} (Springer, New York, 1982).%41

\bibitem{carrol}L.  M.  Pecora  and  T.  L.  Carroll,  \prl{64}{821}{1990}; Phys. Rev. A{\bf 44}, 2374 (1991).
\bibitem{koca}L. Kocarev and U. Parlitz, \prl{74} {5028}{1995}.
\bibitem{ujjwal}S. R. Ujjwal, N. Punetha, R. Ramaswamy, M. Agarwal, and A. Prasad, \chaos{26}{063111}{2016}.%42
\bibitem{maistrenko} L. Larger, B. Penkovsky, and Y. Maistrenko, \prl{111}{054103}{2013}.%43
\bibitem{showalter} S. Nkomo, M. R. Tinsley,  and K. Showalter, \prl{110}{244102}{2013}.%44
\bibitem{camargo}S. Camargo, R. L. Viana, and C. Anteneodo, \pre{85}{036207}{2012}.%45
\bibitem{pecora-chaos} L. M. Pecora, T. L. Carroll, G. A. Johnson, and D. J. Mar, \chaos{\bf{7}}{4}{1997}.%46
\end{thebibliography}
\end{document}